\documentclass[%
 reprint,
 amsmath,amssymb,
 aps,
]{revtex4-2}

\usepackage{graphicx}% Include figure files
\usepackage{dcolumn}% Align table columns on decimal point
\usepackage{bm}% bold math
\usepackage[mathlines]{lineno}% Enable numbering of text and display math
\usepackage{bm}
\usepackage[utf8]{inputenc}
\usepackage{amssymb}
\usepackage{color}
\usepackage{amsmath}
\usepackage{lipsum}
\usepackage{subcaption}
\usepackage{caption}
\begin{document}

\preprint{APS/123-QED}

\title{Sloppy Gear Mechanism for Coupled Stochastic Transportation: from anti-equilibrium flow to infinite selectivity
}% Force line breaks with \\%

\author{Chase Slowey}
\author{Zhiyue Lu}%
 \email{zhiyuelu@unc.edu}
\affiliation{%
 Department of Chemistry, University of North Carolina at Chapel Hill, Chapel Hill, NC 27599-3290}%

\date{\today}

\begin{abstract}
Non-equilibrium transportation of particles through a restricted space (such as porous media or narrow channels) significantly differs from free space. With a simple model of two types of particles competing to transport via a passive single-lane channel connecting two chemical baths, we find two dynamical modes of transportation---the dud mode and the ratchet mode. At the ratchet mode, the gradient flow of one type of particle forces the other into a transient anti-equilibrium flow against its gradient. At the dud mode, the two types of particles both flow according to their gradients. This counter-intuitive non-equilibrium effect can be explained by a sloppy gear mechanism. In addition to the anti-gradient flow observed in the ratchet mode, we find that the non-equilibrium sloppy gear mechanism suppresses the flow of one particle while allowing the flow for the other. This mechanism provides a general theoretical framework to explain and design the selective transportation of particles via passive channels.

\end{abstract}

\maketitle

%\tableofcontents

\section{Introduction}
Systems far from equilibrium are ubiquitous in chemistry, physics and living systems. Their dynamics are typically too far from equilibrium and can not be described by linear response theory or mean-field approaches. Intuitions from near-equilibrium processes can be misleading when directly applied to such non-equilibrium processes. Examples of surprising and even sometimes counter-intuitive non-equilibrium phenomenons includes stochastic pumps \cite{astumian2018stochastic}, kinetic proofreading \cite{hopfield1974kinetic}, chemical reaction oscillations (B-Z reactions) \cite{zhabotinsky1964periodic}, self-organized Rayleigh-Benard patterns \cite{benard1901tourbillons, rayleigh1916lix}, and the Mpemba effect \cite{mpemba1969cool, lu2017nonequilibrium}. The development of stochastic thermodynamics \cite{van2013stochastic} provides us with a powerful set of tools to study dissipative processes far from equilibrium. Here, using a simple kinetic model, we present a counter-intuitive anti-equilibrium phenomenon in the passive transportation of particles through a narrow tube. 

Stochastic transportation of particles via a narrow tube can be well described by the famous set of simple models of exclusion process.~\cite{spitzer5interaction} These models consider a narrow tube as an array of one-dimensional lattice sites in which each site can only be occupied by up to one particle. Particles excludes each other via hard-sphere repulsion and transport through the tube via stochastic hopping between neighboring sites. If the hopping is biased towards one direction, the model becomes the asymmetric exclusion process (or totally asymmetric exclusion process when hopping is only allowed for one direction).~\cite{blythe2006introduction, mallick2015exclusion, schutz1997exact, derrida1998exact, denisov2015totally} These models have successfully captured the essential physical nature of a single type of particle transporting via a restricted space and have demonstrated interesting dynamical phase transitions: when the density of particles in a lattice increase above a threshold, kinetic jamming takes place resulting in a sharp decrease in the transportation rate.\cite{mallick2015exclusion, kim2007dynamic, de2006exact, blythe2006introduction, shaw2003totally, brankov2004totally, derbyshev2012totally, huang2007ramp, nagatani1993shock, sopasakis2006stochastic,song2009totally, antal2000asymmetric, nagatani1995creation, giardina2006direct}

Non-equilibrium transportation can be more complicated if more than one type of particle transport via the same narrow tube. In this situation, the flows of different types of particles can be strongly coupled to each other due to the confinement and demonstrate various counter-intuitive transportation properties. In this work, we report two types of unexpected transportation effects: \emph{anti-equilibrium flow} and \emph{coupling-induced selective transportation}. 
%[We believe that these phenomena can be verified by experiments of gas diffusion via small pores and experiments of granular particles transporting via a narrow tube.]

We build upon existing single-species exclusion process models to study the competitive/cooperative transportation of two types of particles (A and B) across a narrow tube. We model the tube as a symmetric exclusion lattice array of sites coupled to two chemical baths consisting of mixtures of particles A and B. The tube itself is passive, as the internal transportation of particles satisfies detailed balance conditions without an active source of energy, but is driven away from equilibrium by the concentration gradients between the two chemical baths. 
We find that under certain conditions (i.e., varying energy landscapes and concentration gradients for A and B) the tube can generate a counter-equilibrium flow, i.e., one type of particle is temporarily transported from a low concentration bath to a high concentration bath. This counter-intuitive process is not violating the second law of thermodynamics since it is achieved at the cost of entropy production of the other type of particle. This interesting phenomenon indicates that when two types of particles are present, a simple tube acts like a chemical engine (entropy rectifier) and its thermodynamics efficiency analysis resembles those found in the modern models of Maxwell's demons. \cite{maxwell2001theory, mandal2012work, lu2014engineering,lu2019programmable}. 
This counter-intuitive effect can be explained by the strong coupling between the flows of two types of particles, intuitively presented in this work as a \emph{sloppy gear mechanism}.
Moreover, this sloppy gear mechanism demonstrates a surprising selective transportation in tubes. The coupling between the flows of the two species itself could generate a bias, allowing only one type of particle to transport via the tube while blocking the flow of the other, even when the two types of particles are indistinguishable in terms of their interaction with the tube.

The paper is structured as follows. We first introduce a minimal Markov model of two types of particles transporting through a single-track tube, whose non-equilibrium kinetics can be exactly solved by the master equation. In a simple scenario without tuning parameters, we demonstrate a generic anti-equilibrium flow effect where particles appear to spontaneously travel against their gradient (illustrated by an exactly solved phase diagram of ratchet-dud transitions). The phenomenon is explained by an intuitive sloppy gear mechanism. Using the sloppy gear mechanism, we discover that the kinetic coupling between two types of particles (two gears) can be used to predict and optimize the design of highly selective transportation in rather simple tubes. In the end, we treat the tube as a chemical engine and analyze its thermodynamic efficiency.

\section{General kinetic model}
\label{sec:ratchet}

\begin{figure}[h]
\includegraphics[scale = 0.35]{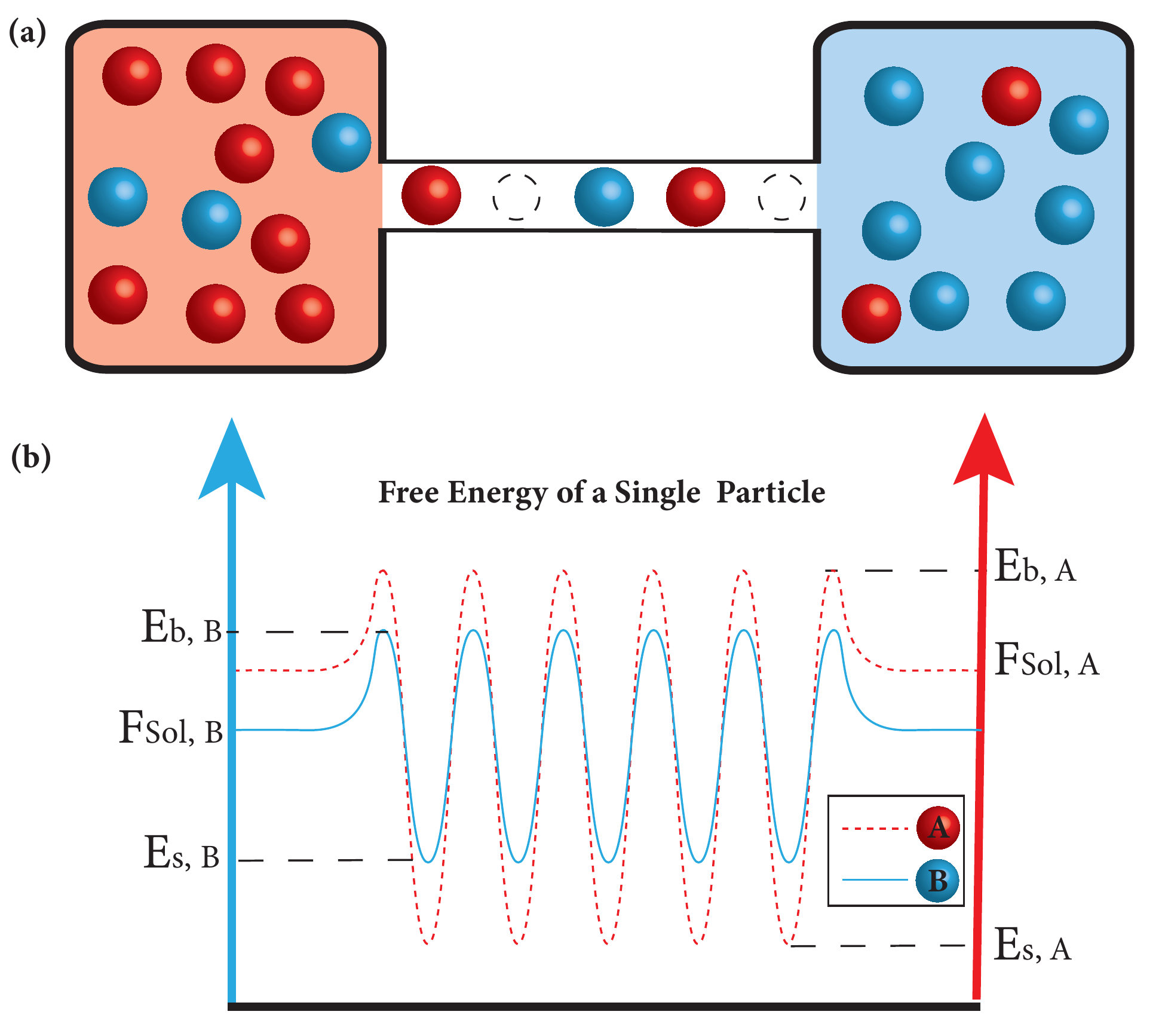}
\centering
\caption{(a) A 5-site single-lane tube connecting two chemical baths of particles A (red) and B (blue). (b) Energy landscape of a single particle. The red dashed line represents the energy landscapes that a particle A experiences at different locations. The 5 energy wells describes the binding energy (affinity) between A and the tube's 5 binding site, $E_{s,A}$. The energy hills $E_{b,A}$ represents the activation barrier for particle A to hop from one site to another, the solvation free energy of a single particle A is represented by $F_{sol,A}$ reflecting As binding affinity with the solvent. For particle B, the landscape is shown as a blue solid line. In the simplest scenario (inert bare-tube), both A and B landscapes are totally flat, when we refer to the tube as an inert tube.}
\label{Fig:Gears}
\end{figure}

Let us describe a general regime of particle transportation via a narrow tube. Consider a 1-D tube with $n$ lattice sites connecting two large chemical baths. These chemical baths consist of mixtures of two types of solute particles (A and B). Both types of particles can enter and exit the edge sites of the tube. We assume that the particles interact with each other via hard-sphere repulsion.
%, and more complicated interactions will be discussed in future works 
Similar to a symmetric exclusion process model,~\cite{oliveira2013mixing, farfan2011hydrostatics, chatterjee2016symmetric} each particle can hop from one lattice site to an empty neighboring site, and each site can hold up to one particle due to the hard-sphere repulsion. The hopping rates of each particle obeys the Arrhenius law~\cite{arrhenius1889reaktionsgeschwindigkeit}, obtained from the energy landscape shown in Fig.~\ref{Fig:Gears}. 

In the general model, we assume that the two types of particles can have distinct binding affinities with the lattice site ($E_{s,X}$), distinct energy barriers for hopping between sites ($E_{b,X}$), and distinct solvation free energies ($F_{sol,X}$) where `X' refers to the particle type A or B. However, for simplicity, we assume that the solvent in both baths are identical and that the $n$ sites of the tube are identical (homogeneous tube). The two chemical baths with different concentrations of particles (different chemical potentials) can lead to a net transportation of particles through the tube. We denote the chemical potential as $\mu_{X,Z} = \beta^{-1} \ln{[X]_Z}$, where $\beta$ is the inverse temperature and $[X]_Z$ is the concentration of particle `X' (i.e., A or B) in bath `Z', where `Z' can represent either the left bath (`L') or the right bath (`R'). The energy that a single particle A (or B) experiences at different locations within or outside the channel is plotted by the energy landscape in Fig.~\ref{Fig:Gears}(b).

At any given concentrations (chemical potentials) the non-equilibrium transportation rate of particles can be calculated by solving for the probability of particle-occupancy configurations (micro-states) of the tube. There are three states for each site: empty, occupied by an A, or occupied by a B. Then, for a 1-D tube that has $n$ sites, there are $N=3^n$ possible configurations (i.e., micro-states) of the tube. For a 5-site tube, there are $3^5=243$ possible micro-states including, but not limited to, (A~A~A~A~A), (A~$\_$~A~$\_$~B), and ($\_~\_~\_~\_~\_$), where ``$\_$'' denotes an empty site. 
The probability over $N$ micro-states is denoted by a $N$-vector $\vec p(t)$ which evolves according to the master equation:
    \begin{equation}
    \label{eqn:master}
        \frac{d \vec p(t)}{dt} =  \hat R \cdot \vec p(t) 
    \end{equation}
where $\hat R$ is a $N\times N$ rate matrix whose off-diagonal element, $R_{ik}$, represents the probability transition rate from micro-state `$k$' to `$i$'. These transitions are among three different types: One particle $X=A~{\rm or}~B$ hops from one site to an empty neighboring site with rate 
\begin{equation}
\label{eq:site-site}
    R_{ik} = \exp[{-\beta (E_{b,X} - E_{s,X})}]~;
\end{equation}
One particle `X' enters an empty left/right edge-site from the left/right chemical bath with rate 
\begin{equation}
    R_{ik} = [X]_{L/R} \cdot \exp[-\beta (E_{b,X} - F_{sol,X})]~;
\end{equation}
One particle of type `X' exits the tube from the left/right and enters the left/right chemical bath with rate
\begin{equation}
    R_{ik} = \exp[- \beta (E_{b,X} - E_{s,X})]~.
\end{equation}
As illustrated in Fig.~\ref{Fig:Gears}, $E_{b,X}$ is the barrier energy for the nearest neighbor site hopping for particle `X' hop, which is assumed to be uniform throughout the entire tube. For convenience, we employ the unit system such that the inverse temperature is unity, $\beta = (K_{B}T)^{-1} = 1$.

\textbf{Separation of time-scale:}  If one considers two well-mixed large baths connected by a microscopic tube, we can separately treat the dynamics within the tube and the dynamics of the baths. On the one hand, the particles enter, leave, and hop within the tube erratically, which occurs at a very fast time scale. On the other hand, the concentrations of the baths are changed by the tube extremely slowly due to the big size of the bath. As a consequence, at any given time we can assume that the tube reaches a non-equilibrium steady state (NESS) with stationary bath concentrations. 
As a result, if one obtains $J_X([\text{A}]_R,[\text{B}]_R)$, the tube's NESS transportation rate of particle `X' at the pseudo-stationary concentrations of the baths ($[\text{A}]_R,[\text{B}]_R$), then one can use such NESS rates to predict the slow change of baths concentrations at the longer time scale. The slow dynamics of the bath concentration follows the following ordinary differential equations (ODEs) 
\begin{equation}
    \dot {[X]}_R= J_X([\text{A}]_R,[\text{B}]_R)
\end{equation}
where $\dot {[X]}_R$ is the time derivative of the concentration of `X' (A or B) in the right bath, and $J_X([\text{A}]_R,[\text{B}]_R)$ is calculated in Eq.~\ref{Eqn: Detailed Prob Current}. Notice that due to the conservation of materials, we only need to specify $[\text{A}]_R,[\text{B}]_R$ to determine the concentration of both particles in the left bath: $[\text{A}]_R+[\text{A}]_L=$~const. and $[\text{B}]_R+[\text{B}]_L=$~const. The ODEs are later illustrated by 2-dim vector fields (normalized) in Fig~\ref{Fig. Relax}.

\textbf{NESS transportation rates:} To obtain the NESS transportation behavior of the tube we solve for the NESS probability distribution $\vec p^{ss}$ of the tube's micro-states, as the null-space of rate matrix:
    \begin{equation}
        \label{eqn: Nullspace}
        \frac{d \vec p^{\,ss}}{dt} =  \hat R \cdot \vec p^{\,ss} = 0 
    \end{equation}
Notice that the probability transition rate matrix, $\hat R$, is parameterized by the pseudo-stationary concentrations of A and B in both baths. 

Using the NESS probability distribution, $\vec p^{\,ss}$, solved for in Eq.~\ref{eqn: Nullspace} with the pseudo-stationary bath concentrations, we can compute the steady state probability transition rates $j^{ss}_{ik}$ for the tube's configuration change from micro-state $k$ to $i$:
\begin{equation}
    \label{eqn: DetailedCurrent}
    j^{ss}_{ik} = R_{ik}  p^{ss}_k
\end{equation}
where $p^{ss}_k$ is the steady-state probability of the tube at micro-state $k$. 
Thus, at the steady state of given bath concentrations, the tube is able to facilitate a net transportation of particle type `X' through the tube. This transportation rate can be computed by first selecting the micro-state transitions that correspond to a particle of type `X' entering the right bath (positive event) and the micro-state transitions that correspond to a particle of type `X' entering the tube from the right bath (negative event). Then by summing the detailed probability currents for all of the events that contributed to the net transportation of particle `X', we can obtain the particle flow rate (transportation rate) $J_{X}$ for particles of type `X' at the NESS for any given bath concentrations:
\begin{equation}
\label{Eqn: Detailed Prob Current}
    J_{X} = \sum_{i,k} j^{ss}_{ik} \cdot T^X_{ik}
\end{equation}
where $ T^{X}_{ik}$ is a current indicator, $ T^{X}_{ik}=1$ for positive events, $ T^{X}_{ik}=-1$ for negative events, and $ T^{X}_{ik}=0$ for transitions that are not associated with particle `X' leaving/entering the right bath.

\textbf{Simplest model: parameter-free inert bare-tube.} The general model described above allows us to calculate non-equilibrium coupled transportation of two types of particles via a narrow $n$-site tube given parameters to tune the interaction between the particle and sites and between particles and the solvent. We emphasize that the novel transportation effects described in this paper can be demonstrated by a simplest parameter-free scenario. In the simplest \emph{parameter-free model}, we set $F_{sol,A} = E_{s,A} = E_{b,A} = F_{sol,B} = E_{s,B} = E_{b,B} = 0$ to represent an \emph{inert bare-tube} with flat energy landscapes. In this case, both types of particles are ``indistinguishable'' to the tube as they interact with the tube's sites identically. Also, since the hopping energy barriers are set to 0, the whole tube can be considered as a smooth bare-tube. Also, due to  $F_{sol,X} =0$, the two types of particles are ``indistinguishable'' in terms of their interaction with the solvent.

\textbf{General model and future extensions:}
The presented model is more general than a bare-tube and thus can be used to study of transportation effects by tuning shape of the energy landscapes. In this case, the two types of particles can assume different interaction intensity with the tube: $E_{s,A} \neq E_{s,B}$, and $E_{b,A} \neq E_{b,B}$, the two types of particles can also have distinct interaction with the solvent: $F_{sol,A}\neq F_{sol,B}$. This general model allows us to design tubes that can facilitate optimized selectivity into the transportation. For future work, this model may be extended to account for the presence of external fields by tilting the energy landscapes, allowing for the examination of charged particles transporting across the tube under an external electrical field. In the tilted case, the model is akin to asymmetric exclusion processes \cite{spitzer1991interaction} with two competing types of particles. Such an extension may be a step toward describing transportation of charged particles such as ions under external electrical field. However, the tilted energy landscape and charge-charge interactions are beyond the scope of this manuscript.

\section{Dud-Ratchet Transition}
\label{Sec. Chem Engine}

\subsection{Anti-equilibrium flow via inert bare-tube}
\begin{figure}
\includegraphics[scale = 0.35]{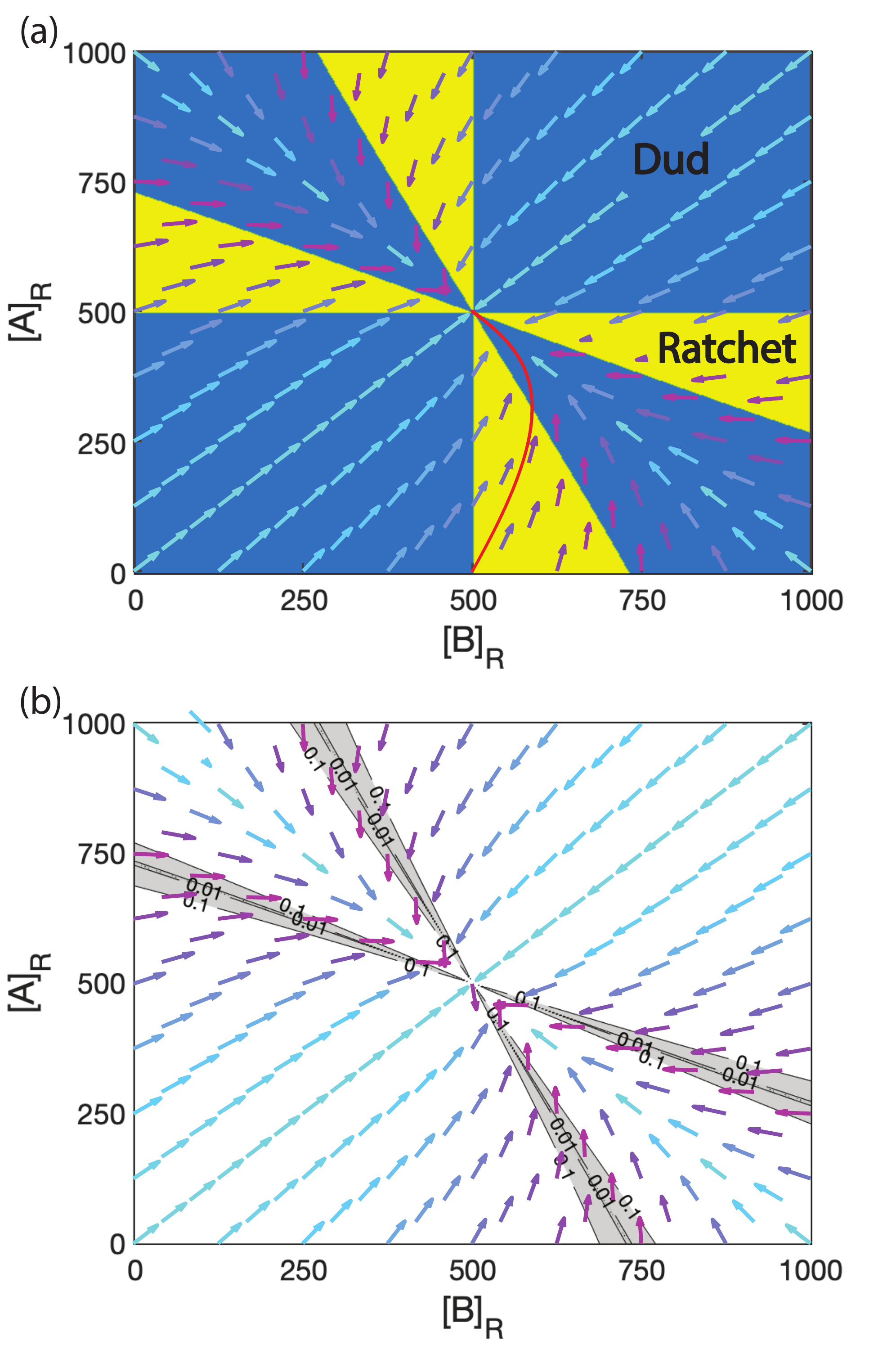}
\centering
\caption{Concentrations' relaxation for both A and B ($J_A$ and $J_B$) illustrated by normalized vector field $\vec f^{\,ss}$ for an inert tube (flat energy landscapes). Each vector represents the normalized NESS flow (normalized rate of concentration change of $[\text{B}]_R$ and $[\text{A}]_R$) at the given bath condition. (a) Two distinct modes of transportation can be found in the system: a ratchet mode (yellow) and dud mode (blue). At the ratchet mode, the rate vector points away from the final equilibrium (500,500), indicating an anti-equilibrium flow. As a consequence of the ratchet mode, if the bath starts at (500,0), it follows the red line, first increasing the concentration gradient for particle B before final equilibrium is reached, contrary to equilibrium intuition where both concentrations monotonically relax to (500,500). (b) On the same vector field, we illustrate the contour lines for the tube's selectivity ($\xi$). Shaded regions indicate where the tube is highly selective $\xi\leq 0.1$. Arrows in both figures are colored according to the selectivity $\xi$.}
\label{Fig. Relax}
\end{figure}

Here we demonstrate that an anti-equilibrium flow can be observed via narrow tubes. First let us examine the simplest case -- an inert 5-site bare-tube, where the two types of particles, A and B, are equally inert to the tube's binding sites and inert to the solvents. \footnote{Notice that here we choose a 5-site tube for illustrative purposes. We demonstrate this effect in a short tube to avoid confusing this novel effect with the dynamical phase transitions discovered in the TASEP or ASEP models where the transitions occurs at the long-tube limit.}  In this case, the energy landscapes  for both particle A and particle B are perfectly flat horizontal lines (simpler than that in Fig.~\ref{Fig:Gears}). Specifically in the simple model we set $E_{b,X} = F_{sol,X} = E_{s,X} = 0$ for both types of particles. Under this assumption, by numerically solving for Eq.~\ref{Eqn: Detailed Prob Current}, the NESS current of particle `X', $J_X([\text{A}]_R,[\text{B}]_R)$, of a 5-site tube at all possible combinations of concentrations under the restriction that $[\text{A}]_R +[\text{A}]_L = 1000$ and $[\text{B}]_R +[\text{B}]_L = 1000$, we can represent the concentration evolution by a normalized vector field, 
\begin{equation}
   \vec f^{\,ss} =\frac{(\dot{[\text{B}]}_R,\dot{[\text{A}]}_R)}{ |(\dot{[\text{B}]}_R,\dot{[\text{A}]}_R)|}=\frac{(J_B,J_A)}{ |(J_B,J_A)|}
\end{equation} 
which is a function of the baths' concentrations, illustrated as unit arrows  Fig.~\ref{Fig. Relax}(a). 

In this simplest parameter-free model, one can identify a transient anti-equilibrium transportation. Starting at an initial concentrations $[\text{B}]_R=[\text{B}]_L=500$, $[\text{A}]_R=0$, and $[\text{A}]_L=1000$, one can observe that its relaxation toward final equilibrium is non-monotonic. The combined relaxation of A and B is illustrated in Fig.~\ref{Fig. Relax}(a) by the red trajectory tracing the vector field starting from (500,0) and ending at (500,500). The anti-equilibrium flow for B is found at the begining of the relaxation: even though B particles are initiated at thermal equilibrium ($[\text{B}]_R=[\text{B}]_L=500$), a spontaneous anti-equilibrium flow of B toward the right bath is observed Fig.~\ref{Fig. Relax}(a). Following the trajectory in the concentration space, this anti-equilibrium flow of B starting at (500,0) temporarily increases  $[\text{B}]_R$ to around $570$ before eventually relaxing to the ultimate thermal equilibrium ($[\text{A}]_R=[\text{A}]_L=[\text{B}]_R=[\text{B}]_L=500$). This anti-equilibrium flow of B appears to violate the second law of thermodynamics, as the entropy of particle B spontaneously decreases. However, this anti-equilibrium transportation is driven by the spontaneous process of the coupled flow of particle A and the total entropy of the system remains increasing in time. This counter-intuitive dynamics indicate that even a simple tube is able to function as a chemical engine, rectifying the chemical potential gradient of particle A to replenish the chemical potential gradient for particle B, which will be discussed in detail in Sec.~\ref{Sec:Efficiency}. In this section, we focus on two distinct operating modes of the tube. Notice that the anti-equilibrium flow, even though is demonstrated first by a simple bare-tube, is a general effect for more complicated energy landscapes. Below we provide a general description and a kinetic physical explanation for this effect.

\subsection{Sloppy gear mechanism} 
\begin{figure}[h]
\includegraphics[scale = 0.35]{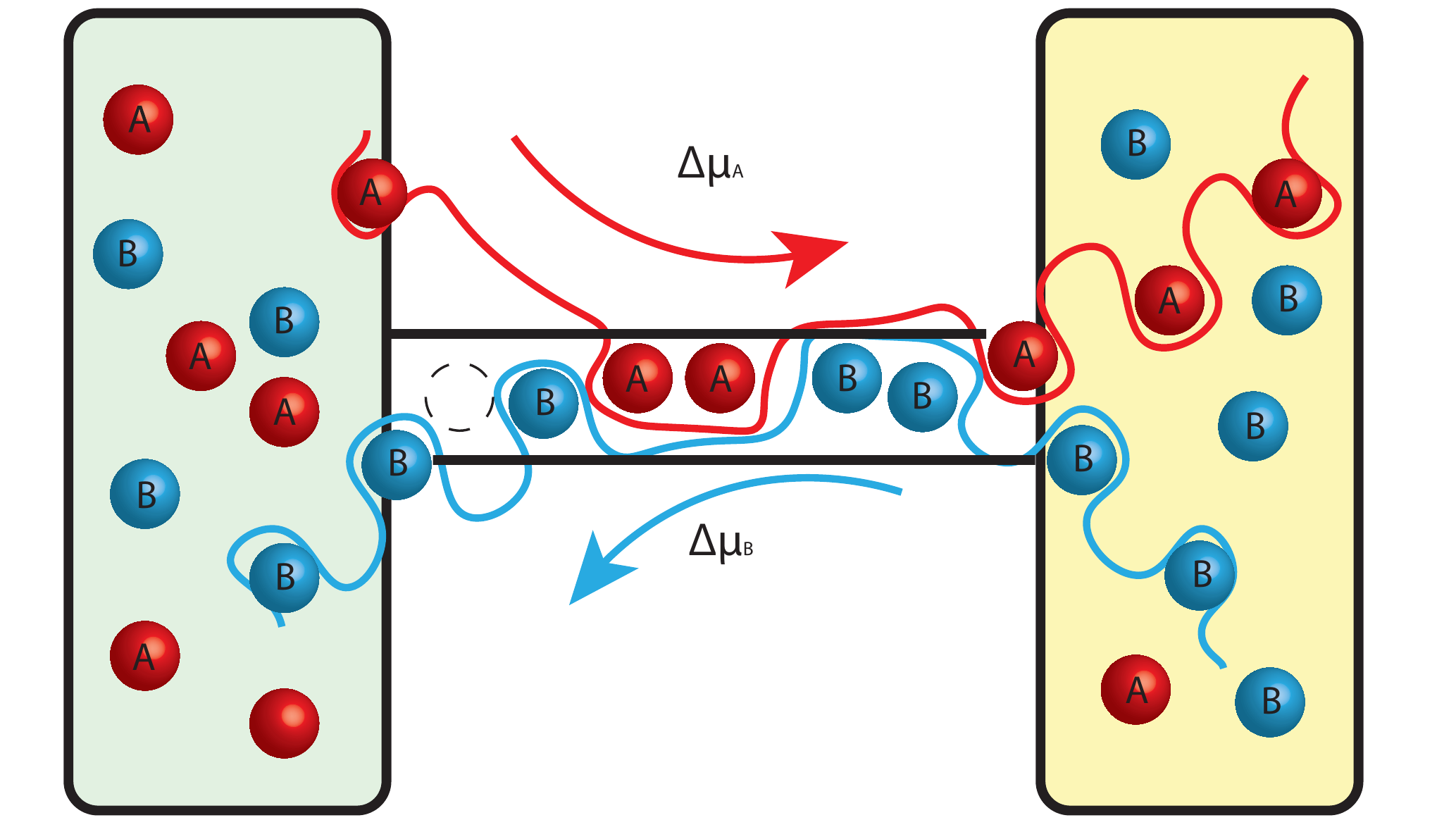}
\centering

\caption{Cartoon illustration of the proposed sloppy gear mechanism. Each particle, as it enters the tube, mimics a tooth of an imagined gear (A in red or B in blue). The gear teeth are loose as their relative positions can change due to allowed particle hopping. Both A and B imaginary gears are driven by their concentration gradients ($\Delta \mu_A$ and $\Delta \mu_B$). Due to the stochastic nature of the particles in the tube, the gear teeth are erratic and sloppy and gears may slip past each other. The longer the tube, and the more particles are found in the tube, the stronger the coupling between two imaginary gears. The strong coupling between two gears results in the ratchet mode and can lead to a highly selective transportation of particles.}
\label{Fig: Sloppy Gear}

\end{figure}

Using the vector field in Fig.~\ref{Fig. Relax}(a), we identify two distinct kinetic modes of transportation---the \emph{ratchet mode} and the \emph{dud mode}. At a given concentration (i.e., one point in Fig.~\ref{Fig. Relax}(a)), if one type of particle is transported against its own concentration gradient (i.e., $\vec f^{\,ss}$'s projection on that particle's axis points away from the center), the tube facilities an anti-equilibrium flow and the tube is in the ratchet mode (yellow shade). If both types of particles are transported down their concentration gradient (i.e., $\vec f^{\,ss}$'s projections to the both axis both point toward the center), the tube is in the dud mode (blue shade). The ratchet mode, where one type of particle is transported against its own gradient can be explained by a non-equilibrium kinetic mechanism---the \emph{sloppy gear mechanism}, illustrated in Fig.~\ref{Fig: Sloppy Gear} and detailed below. 

The sloppy gear mechanism explains the counter-intuitive ratchet mode by the stochastic coupling between the particle flows of A and B. Fig.~\ref{Fig: Sloppy Gear}associates the transportation behavior with the stochastic dynamics of the multi-site particle occupancy configurations, e.g., ($~\_$~B~A~A~B~B~A) sketched for a 7-site tube in Fig.~\ref{Fig: Sloppy Gear}. The site number 7 is chosen arbitrarily for illustrative purposes. Suppose the gradient-driven-flows of particles A and B can be loosely visualized as two imaginary sloppy gears, A and B. The imaginary teeth of each gear represent the particles A or B within the tube. The two gears are loosely coupled since both the number of gear teeth in the tube and the positions of the teeth are stochastic. However, due to the exclusion between particles, the teeth of the gear (particles) cannot bypass each other, resulting in a gear-gear coupling. Consider $[\text{A}]_L>[\text{A}]_R$ and $[\text{B}]_L<[\text{B}]_R$, then the imaginary gear teeth of A (or B) favors motion to the right (or left). Then, the two imaginary gears push against each other via teeth-teeth exclusion interactions. When there are enough teeth in the tube for both particles, the motion of the two imaginary gears are strongly coupled (i.e., the motion of two types of particles are strongly correlated), and the flow is dominated by the gear with the ``stronger'' driving force. In this scenario, the stronger gear drives the other against gradient force, resulting in a ratchet mode of transportation, where one type of particle flows against its own gradient. However, when the number of teeth in the tube is low, the two imaginary gears are too sloppy to slip against each other, allowing for both types of particles to transport down their gradients, as seen in the dud mode. In the dud mode, the correlation between the flow of the two types of particles is not strong enough to invert the flow's direction but is still present as discussed in the section of selectivity.

\begin{figure}[h]
\centering
\includegraphics[scale = 0.35]{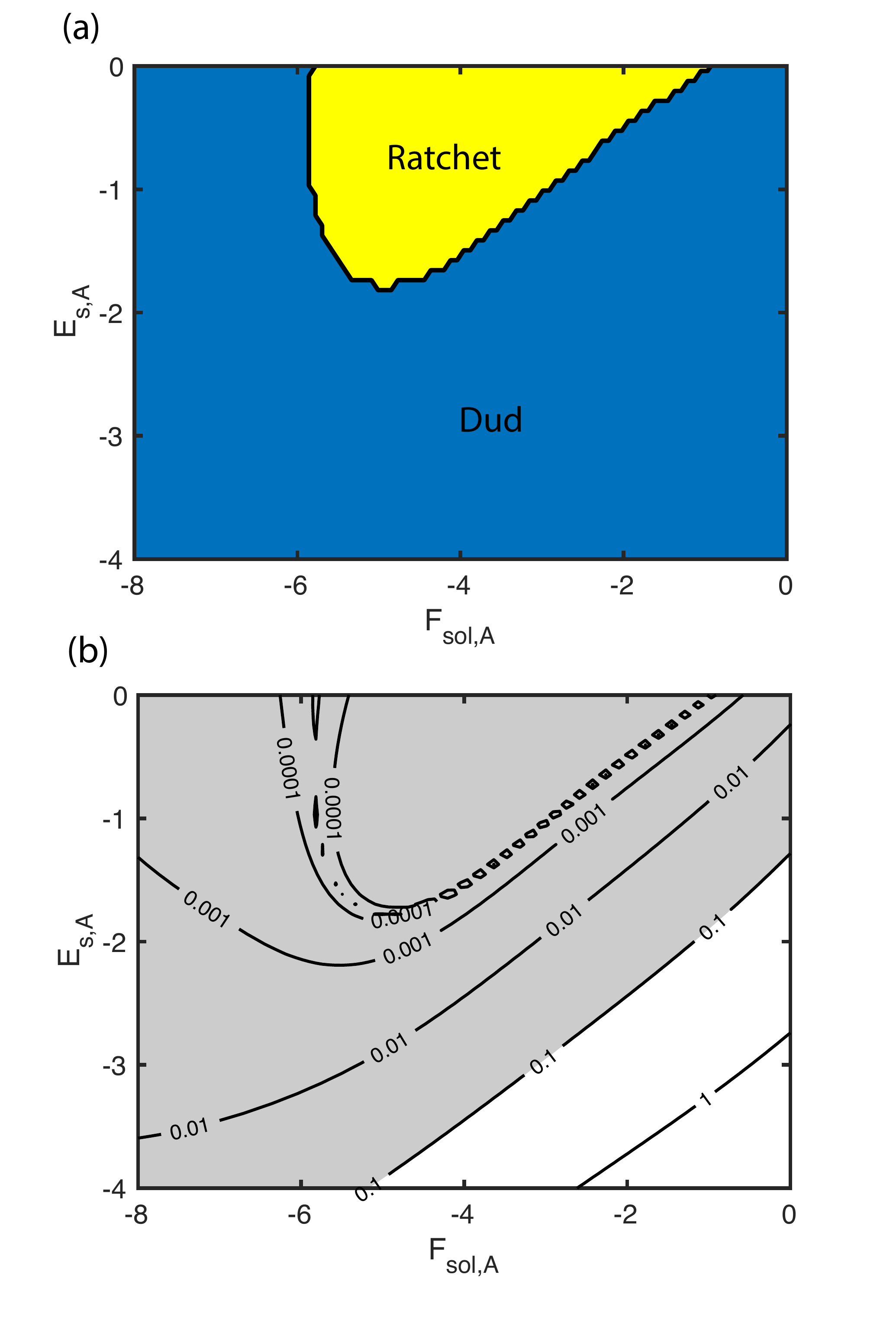}
\caption{At a a static concentration $[\text{A}]_L = 1$, $[\text{A}]_R = [\text{B}]_L = 1000$, and $[\text{B}]_R = 100$, NESS behavior of a non-inert tube with non-flat energy landscape ($E_{s,B} = F_{sol,B} = -4.0$, $E_{b,B} = -1.2$, $E_{b,A} = -2.2$. (a) The ratchet and dud modes of transportation can be achievable by tuning the interaction energy ($E_{s,A}$) and solvation free energy ($F_{sol,A}$) for A particle. In the ratchet mode B particles are forced to go against their gradient due to the flow of A. (b) Illustration of tube's selectivity ($\xi$). The shaded regions ($\xi\leq 0.1$) indicate where the selective tube strongly favors the current of A and suppresses the current of B. Notice that selective transportation $\xi\leq 0.1$ can be achieved beyond the ratchet mode.}
% In the `ratchet' phase, the current of particle B flows from low concentration to high concentration, driven by a ratchet effect induced by the flow of A. In the `dud' phase, both particle A and B flow from their high concentration side to their low concentration side.}
\label{Fig: PhaseTransition}
\end{figure}

The two kinetic modes of operations discussed above are not restricted to inert bare-tubes (flat energy landscapes). Let us consider a general case where the two types of particles have different interaction affinities to the tube's sites and different solvation free energies in the solvent: We arbitrarily choose the energy landscape such that $E_{s,B} = F_{sol,B} = -4.0$, $E_{b,B} = -1.2$, $E_{b,A} = -2.2$, while allowing for As interaction with the tube and the bath to vary. Thus $E_{s,A}$ and $F_{sol,A}$ are the only two free parameters from which we obtain a 2-dim phase diagram. We numerically solve for the NESS currents of the specified 5-site tube connecting two baths with fixed concentrations, $[\text{A}]_L = 1$, $[\text{A}]_R = [\text{B}]_L = 1000$, and $[\text{B}]_R = 100$, and confirm that we can find both the ratchet mode and the dud mode in the $E_{s,A}$--$F_{sol,A}$ diagram, illustrated in Fig.~\ref{Fig: PhaseTransition}(a).

Notice that the ratchet mode where particle A inverts the flow of particle B cannot be simply explained merely by the particle-tube interaction ($E_{s,A}$) nor the particle-solvent interaction ($F_{sol,A}$). Otherwise, one should not observe this effect in the inert bare-tube. However, the sloppy gear mechanism (Fig.~\ref{Fig: Sloppy Gear}), is a non-equilibrium kinetic theory, which explains the anti-equilibrium transportation based on the kinetic coupling between the two types of particles within the narrow tube. 
To correctly understand the roles played by the particle-tube interaction ($E_{s,A}$) and particle-solvent interaction ($F_{sol,A}$) in the transportation effect, one should focus their impact on the stochastic occupancy configurations (micro-states) of the tube. For example, altering the energy landscape non-trivially changes the typical ``teeth arrangements'' of the two sloppy gears, which results in a complex non-monotonic phase boundary between the ratchet mode and the dud mode shown in Fig.~\ref{Fig: PhaseTransition}(a).  

\section{Selective Transportation}
\label{Sec: Selectivity}
\subsection{Sloppy gear induced selectivity}
The sloppy gear mechanism reveals that the non-equilibrium coupling between the flows of two types of particles  can result in selective transportation. Due to the sloppy gear teeth coupling (particle-particle exclusion), one type of particle's flow (the dominant imaginary gear's teeth) can suppress the flow of the other, resulting in the tube strongly favoring the transportation of one type of particle while blocking the flow of the other. Here we name the type of particle whose transportation current is suppressed the ``unfavored'' type, and the other type, ``favored''. Then we can denote the selectivity of the tube by the absolute value of the transportation current ratio between the unfavored (`uf') particle and the favored (`f') particle:
\begin{equation}
    \xi =\left | \frac{J_\text{uf}}{J_\text{f}}\right |
\end{equation}
where the current $J_\text{uf}$ and $J_\text{f}$ refers to the NESS currents of the unfavored type of particle and favored type of particle, as defined in Eqn. \ref{Eqn: Detailed Prob Current}. For illustrative purposes, let us define a threshold for ``high selectivity" as $\xi\leq 0.1$. This allows us to illustrate the highly selective range of parameters as shaded areas on the selectivity contour diagrams where we have shown the contour plots of $\xi$ in Fig.~\ref{Fig. Relax}(b) and Fig.~\ref{Fig: PhaseTransition}(b). Notice that the tube can be \emph{infinitely selective} ($\xi=0$) at the phase boundary between the ratchet mode and the dud mode, as the NESS current of the unfavored particle is totally stalled by the sloppy gear interaction with the favored particle flow. Also notice that the tube is highly selective not only in the ratchet phase (see shaded area in Fig.~\ref{Fig: PhaseTransition}b).

It is worth to mention that the selective transportation is a non-equilibrium effect caused by the coupling between the two particle flows (sloppy gear mechanism). Thus it cannot be explained by intuition from equilibrium thermodynamics. Intuitively, the selectivity of transportation of particles can be determined by the particles' interaction with tube's binding sites ($E_{s,X}$), particle-solvent interactions ($F_{sol,X}$), and the hopping barrier between neighboring sites ($E_{b,X}$). However, as shown in Fig.~\ref{Fig. Relax}(b), even when the two types of particles are physically and chemically indistinguishable in terms of their interactions with the tube and the solvents, the coupling between their flows can result in an infinitely high selectivity, where one type of particle with a stronger chemical potential gradient can dominate the tube while suppressing the flow of the other type to 0. In this example, the selectivity is achieved purely by the coupling between the non-equilibrium flows. Thus, in the simplest model, the equilibrium distinctions between particles A and B are totally removed from the discussion of selective transportation (e.g., at equilibrium, particle A and B are indistinguishable to the tube and to the baths, yet the tube can still demonstrate selective transportation). 

This sloppy gear mechanism for tube selectivity provides a novel non-equilibrium perspective to understand and optimize selective transportation in narrow tubes. Rather than focusing on the equilibrium distinctions between A and B (e.g., $E_{s,X}$, $E_{b,X}$, and $F_{sol,X}$), selective transportation can be understood by flow coupling between the two species. The strength of the flow for each species are impacted by the kinetic rates of particle-hopping, which are influenced by concentration gradients, temperature, $E_{s,X}$, $E_{b,X}$, $F_{sol,X}$, and the kinetic micro-state of the tube. Below, we demonstrate that the sloppy gear mechanism can be used to design highly selective transportation tubes. 

\subsection{Optimal design of selectivity} 
\begin{figure}[h]
\centering
  \includegraphics[scale = 0.40]{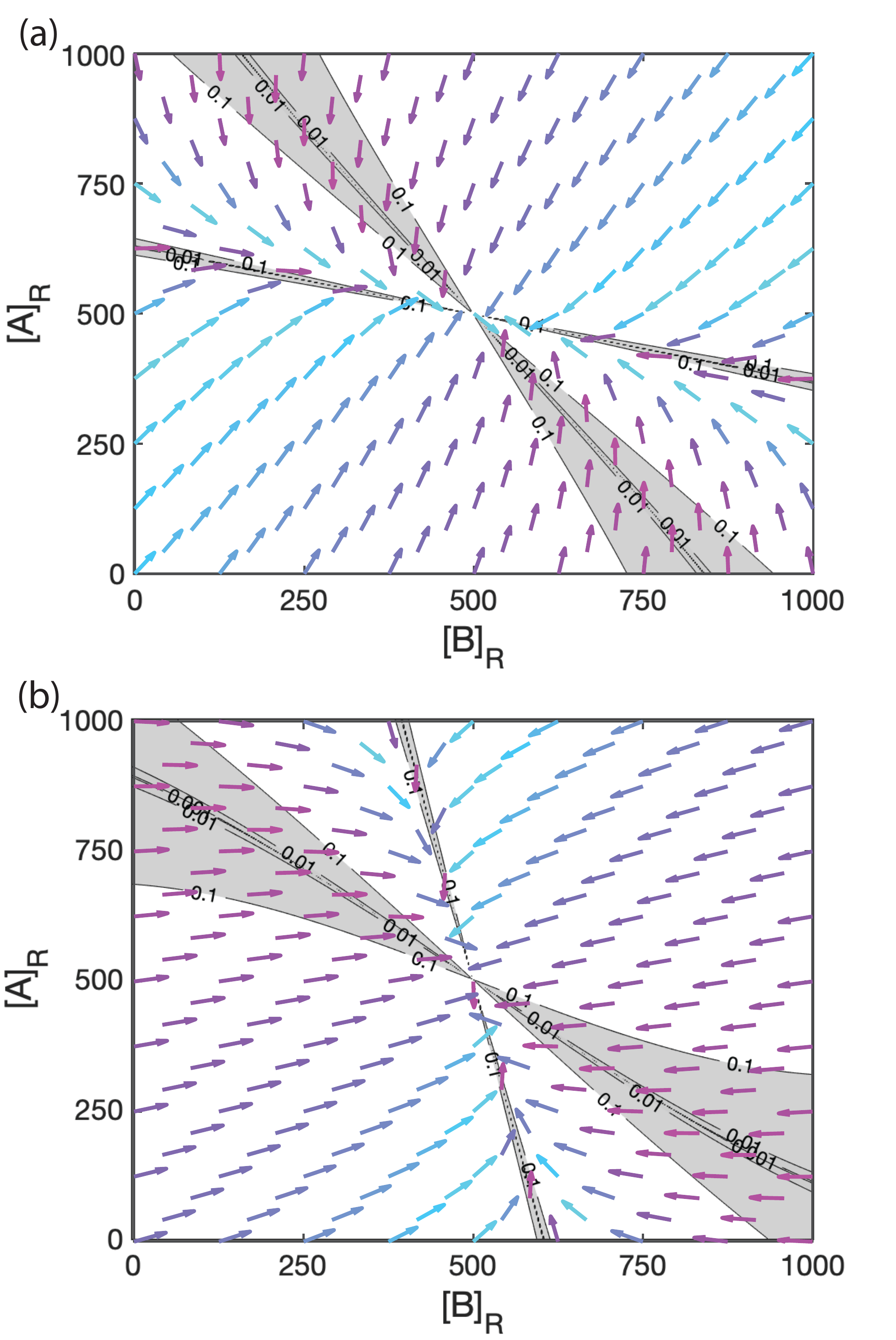}
\caption{Comparison of the selective transportation for two types of landscapes. Shown side-by-side are normalized vector fields $\vec f^{\,ss}$'s for modified inert tubes with two different alternations: (a) $E_{s,A} = -1$ and (b) $F_{sol,A} = -1$. Shaded regions indicate high selectivity $\xi\leq 0.1$. Compared to a purely inert tube (Fig.~\ref{Fig. Relax}(b)), both alternations results in an increased selectivity. Notice that the high selectivity region is larger in (b) than in (a), indicating the selectivity is more sensitive to particles' difference in solvation free energies. }
\label{fig: SiteSelectivity}
\end{figure}

Above discussion have demonstrated that even an inert bare-tube can selectively transport particles, which is illustrated by Fig.~\ref{Fig. Relax}(b), where $E_{b,A} = F_{sol,A} = E_{s,A} = 0$ and $E_{b,B} = F_{sol,B} = E_{s,B} = 0$. However, the highly selective regions are relatively small compared to the whole phase diagram, and it is obviously not optimized in terms of selectivity. 
Here we explore the optimal design of selective transportation for the general model of a narrow tube (i.e., with non-flat energy landscapes caused by chemical decorations at each site). Specifically, let us consider the particle--site interaction and particle--solvent interactions and study their effect on both the current and the selectivity. In contrast to Fig.~\ref{Fig. Relax}(b) obtained for an inert bare-tube, we observe enhanced selectivity as we slightly alter the bare-tube into realistic tubes (see the larger area of highly selective region in Fig.~\ref{fig: SiteSelectivity}). Here we numerically calculate the selectivity contour for modified inert tubes (modifying from $E_{b,X} = F_{sol,X} = E_{s,X} = 0$ for both `X' = A and B) with the only modification that (1) $E_{s,A} = -1$, resulting in an enhanced selectivity favoring the flow of A shown in Fig.~\ref{fig: SiteSelectivity}(a), or the only modification that (2) $F_{sol,A} = -1$, resulting an enhanced selectivity favoring the flow of B shown in Fig.~\ref{fig: SiteSelectivity}(b). 
%This result agrees with the intuition obtained in the study of real ion channels.~\cite{bezanilla1972negative, doyle1998structure, gouaux2005principles, fowler2008selectivity}

As a summary, the sloppy gear mechanism and the kinetic analysis provides us with a guiding principle toward designing selective transportation in rather simple models of narrow tubes. We find that both $E_{s,A}$, the particle's interaction with the tube and $F_{sol,A}$, the particle's interaction with the solvent can significantly and non-monotonically impact the selectivity. By comparing Fig.~\ref{fig: SiteSelectivity}(a) and (b), we notice that altering $F_{sol,A}$, particle-bath interaction can more prominently impact the selectivity of the tube. Moreover, through a simple kinetic argument, we expect that altering $F_{sol,A}$ is advantageous over altering $E_{s,A}$ toward designing a \emph{low-resistance selective tube}: reducing $E_{s,A}$ can indeed make the sloppy gear A dominant over the gear B, and favor the flow of particle A over B. However, decreasing $E_{s,A}$ can significantly reduce the stochastic hopping rate of particle A (see Eq.~\ref{eq:site-site}), and thus also reduce the transportation current of particle A. In contrast, via increasing $F_{sol,A}$ or decreasing $F_{sol,B}$, we can achieve the same enhancement favoring the flow of particle A because it increases the probability for particle As to occupy the tube, thus making the `gear' A the dominant gear. Moreover, this enhancement is not at the cost of slowing down the particle hopping rates for either A or B because it does not impact Eq.~\ref{eq:site-site}. Thus if one intends to enhance the selectivity without impeding the flow rate of the favored particle, the design should lean more toward altering $F_{sol,A}$ rather than altering $E_{s,A}$. \footnote{One may argue that similar optimized performance of selective transportation of A could be achieved via increasing $E_{s,B}$, i.e.,introducing a strong repulsion between the particle B and the tube. However, this solution is trivial as it is simply requires that one find a tube that is simply not permeable to particle B. 
%Such a trivial design cannot explain the strange selectivity in KcsA ion channels which allows big particles K$^+$to transport through but simultaneously blocks smaller particles Na$^+$.
}

\section{Entropic Efficiency}
\label{Sec:Efficiency}
At the ratchet mode, the tube can be considered as a chemical engine, rectifying chemical energy of one type of particle and using it to increase the chemical potential difference in another type of particle. Here let us examine the thermodynamic entropy production. For simplicity, let us consider a ratchet mode where A pushes 'B' and thus forces B to an anti-equilibrium flow against its own gradient. As a consequence, the total entropy of particle B decreases over time. (Similar arguments can be made for the other ratchet mode where particle B pushes A.) Thus, we have a mechanism to harness the spontaneous increased entropy of particle A and use it to reduce the entropy for particle B. At the NESS, the rate of entropy decrease of the particle B can be computed as
\begin{equation}
\label{Eqn: Entropy Production}
    \dot S_B = J_B \ln{\frac{[\text{B}]_L}{[\text{B}]_R}} 
\end{equation}
The second law of thermodynamics requires that the total entropy for both particle A and B must not decrease in time:
\begin{equation}
\label{Eqn:sec-law}
    \dot S_{tot} = \dot S_{A} + \dot S_{B} \geq 0
\end{equation} 
Thus, we characterize the efficiency of this chemical engine (in the ratchet mode) by the steady-state entropy efficiency which must not exceed $1$ according to the second law of thermodynamics:
\begin{equation}
    \label{Eqn: efficiency}
    \eta_B = \frac{-\dot S_B }{\dot S_A} = -\frac{J_B}{J_A}\cdot \frac{\mu_{B,R} - \mu_{B,L}}{\mu_{A,R} - \mu_{A,L}} \leq 1
\end{equation}
If one chooses to study the ratchet effect where particle B pushes A, then efficiency should be defined as $\eta_A = \frac{-\dot S_A }{\dot S_B}$. Our result confirms that the entropic efficiency is always below $1$. 

\begin{figure}[h]
\centering
\includegraphics[scale= .35]{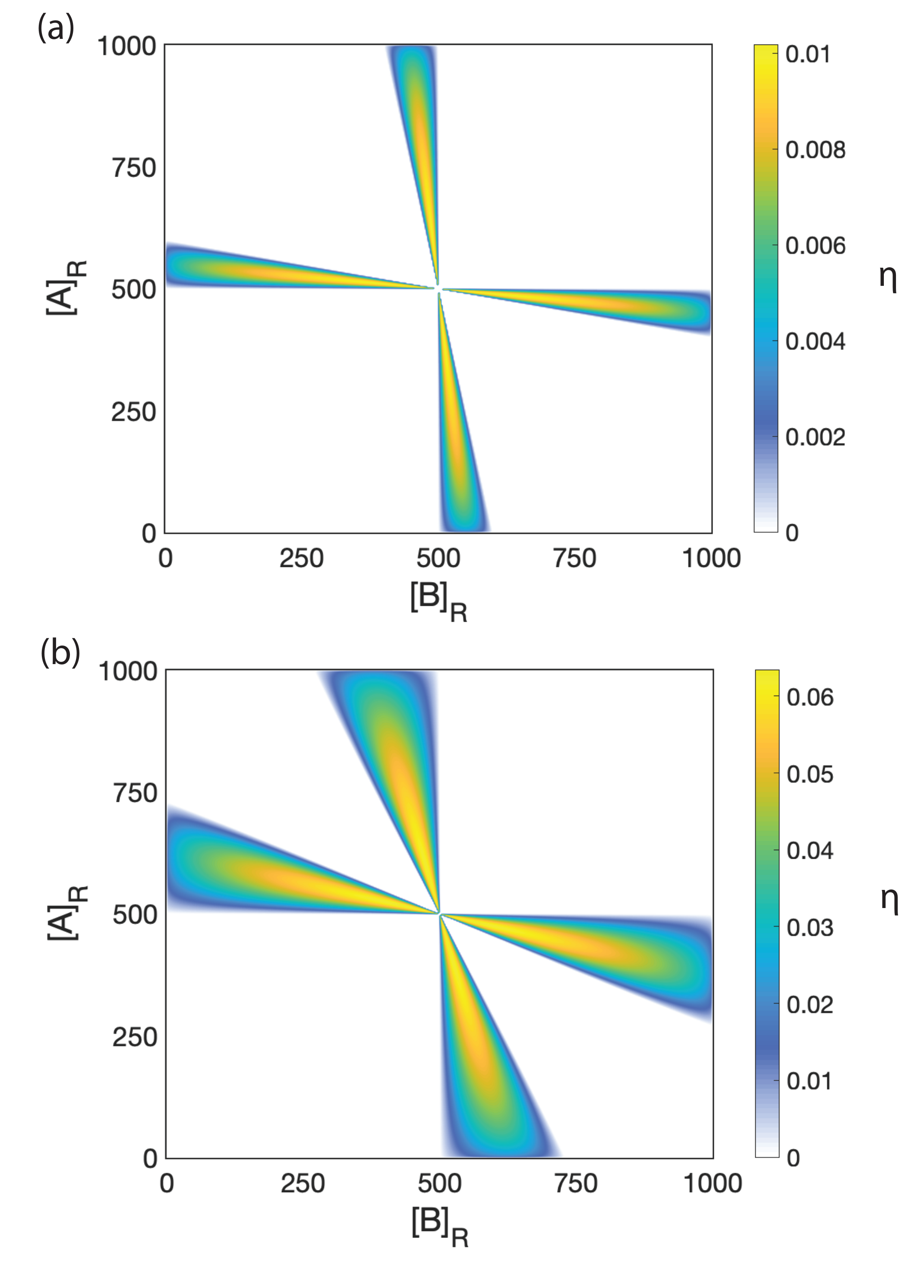}
\caption{Ratchet efficiency, $\eta$, for inert tubes (flat energy landscapes) of different length: (a) 3-site tube and (b) 5-site tube. In agreement with intuition from the sloppy gear mechanism, the longer the tube, the larger the ratchet mode region. Also, the longer the tube, the higher the maximum efficiency, as the two gears are less likely to slip through.}
\label{fig:Eff}
\end{figure}

Here we briefly discuss the efficiency of the ratchet modes obtained at different tube lengths, see Fig.~\ref{fig:Eff}. Our results indicate that the ratchet effect is more prominent in longer tubes, accompanied by a higher maximum entropic efficiency. This can be explained by the sloppy gear mechanism: As the tube becomes longer the two imaginary gears have more possible teeth biting with each other and thus strengthen the kinetic coupling between the flows of particles A and B, reducing the possibility for the two gears to slip by one another (suppression of the dud mode). The longer the tube, the greater the coupling between the flows for particles A and B. It should be noted that, as expected, the efficiency of the ratchet effect reaches its maximum near the ``linear-response regime'' where concentration gradients for both particle types is near zero. Similar efficiency maps with an engine mode and dud mode have been found in various designs of stochastic information engines or modern variations of Maxwell's demons. \cite{mandal2012work, lu2014engineering, lu2019programmable}\\

\section{Conclusion}
This work demonstrates a generic anti-equilibrium transportation phenomenon via a simple SEP-like model describing two types of particles transporting through a purely passive narrow tube. The anti-equilibrium transportation (demonstrated in a bare-tube and then more complex tubes) can be explained by an intuitive sloppy gear mechanism and this general effect can occur without carefully tuning the interactions between the particles and the tube/baths. Moreover, this effect sheds light on a possible mechanism to explain the selective transportation via narrow tubes. By using this simple model we seek the optimal design of selective transportation via narrow tubes. Although both the particle-tube interaction $E_{s,X}$ and the particle-bath interaction $F_{sol,X}$ alter the selectivity of the tube, their influences on the performance of the selective tube are not equivalent: Intuitively, decreasing $E_{s,A}$ or increasing $F_{sol,A}$ would both favor the transportation of particle $A$ over $B$. However, decreasing $E_{s,A}$ reduces the Arrhenius transportation rate and comes with a cost of impeding the transportation rate of $A$. Whereas increasing $F_{sol,A}$ (or decreasing $F_{sol,B}$) does not impede the transportation rate of $A$ within the tube. Thus the optimal design of the selective tube should be achieved via engineering the particle-solvation interaction.

From a thermodynamic perspective, the tube functioning at the ratchet mode behaves like a free energy transducer, harnessing the free energy decrease from one type of particle (that transports down the concentration gradient) and storing it into the free energy of the other type of particle (increasing concentration gradient). The analysis of the entropic efficiency shows that the longer the tube, the more efficient the free energy transduction. 

%Numerical codes used to solve for the model is available at:  \url{https://github.com/Slowey15/ParticleChannel}

\section{Acknowledgments}
This work is inspired from discussion with Michael Zwolak, Subin Sahu, at the workshop ``Information Engines at the Frontiers of Nanoscale Thermodynamics'' at TSRC, in 2019. The authors also appreciate Max Berkowitz, Hong Qian, Chris Jarzynski, Sa Hoon Min, and Yueheng Lan for useful discussions. This work is supported by the University start-up fund at UNC-Chapel Hill. We would like to thank the University of North Carolina at Chapel Hill and the Research Computing group for providing computational resources and support that have contributed to these research results. 

\bibliography{apssamp}% Produces the bibliography via BibTeX.

\end{document}